# Continuous Wavelet Vocoder-based Decomposition of Parametric Speech Waveform Synthesis

*Mohammed Salah Al-Radhi[1], Tamás Gábor Csapó[1,2], Csaba Zainkó[1], Géza Németh[1]*

[1]Department of Telecommunications and Media Informatics
Budapest University of Technology and Economics, Budapest, Hungary
[2]MTA-ELTE Lendület Lingual Articulation Research Group, Budapest, Hungary
`{malradhi,csapot,zainko,nemeth}@tmit.bme.hu`

## Abstract

To date, various speech technology systems have adopted the vocoder approach, a method for synthesizing speech waveform that shows a major role in the performance of statistical parametric speech synthesis. However, conventional source-filter systems (i.e., STRAIGHT) and sinusoidal models (i.e., MagPhase) tend to produce over-smoothed spectra, which often result in muffled and buzzy synthesized text-to-speech (TTS). WaveNet, one of the best models that nearly resembles the human voice, has to generate a waveform in a time-consuming sequential manner with an extremely complex structure of its neural networks. WaveNet needs large quantities of voice data before accurate predictions can be obtained. In order to motivate a new, alternative approach to these issues, we present an updated synthesizer, which is a simple signal model to train and easy to generate waveforms, using Continuous Wavelet Transform (CWT) to characterize and decompose speech features. CWT provides time and frequency resolutions different from those of the short-time Fourier transform. It can also retain the fine spectral envelope and achieve high controllability of the structure closer to human auditory scales. We confirmed through experiments that our speech synthesis system was able to provide natural-sounding synthetic speech and outperformed the state-of-the-art WaveNet vocoder.

## 1. Introduction

The application areas of speech synthesis are expanding into new domains, from newsreaders and simple task-oriented dialogues to multi-turn conversations. Text-to-speech (TTS) synthesis is an essential technology for users and computers to engage in natural spoken dialogue. In this regard, it is desirable to have vocoders synthesize high-quality voices from the speech parameters and be robust to artificial changes. So that vocoder-based TTS systems can synthesize stable speech under the parameters not included in the training data, and that users can process the synthetic speech according to their preference.

Several end-to-end neural models such as WaveNet [1], WaveGlow [2], and LPCNet [3] have been proposed recently and can synthesize high-fidelity speech in TTS [4]. These models are frequently used as vocoders to convert acoustic features, e.g., the spectrogram into speech waveforms. However, in the current development and production scenario, it is important not only to achieve full-band and high-quality synthesis but also to allow users to control speech characteristics according to their preferences. Statistical parametric speech synthesis (SPSS) [5] is applied to be a full-band and highly-controllable TTS system. It uses acoustic features as a low-dimensional intermediate representation for generating the speech waveform from the text. In SPSS, those features significantly influence the quality of synthetic speech and controllability.

Most modern speech synthesis and recognition systems use either perceptual linear prediction or cepstral coefficients as acoustic features. Mel-cepstrum [6] is a well-known example of representation; it approximates the spectral envelope with trigonometric functions' superposition. However, statistical averaging of mel-cepstrum in SPSS changes the entire original structure and significantly degrades synthetic speech quality. In this regard, conventional TTS systems using cepstral features tend to produce over smoothed spectra, which often result in muffled and buzzy synthesized speech. Although some attempts were made to emphasize the peaks and dips of generated spectra via post-processing [7], it is generally difficult to restore original peaks and dips once spectra are over-smoothed. Also, decomposition by trigonometric functions does not result in high controllability. To address this problem, approximation of speech parameters in an alternative speech representation based on wavelet transform is proposed in this paper. On this basis we formulate a new generative model for TTS.

The wavelet transform [8] is becoming a common tool for analyzing localized variations of power in both time and frequency domains. It has been previously used in a variety of applications in speech processing; these include speech enhancement [9], speech segmentation and classification [10], pitch detection and voice conversion [11]. Several studies have been conducted to determine an effective and practical method for controlling the speech synthesis voice. The composite wavelet model was proposed in [12] as an alternative of the vocoder (regarded as convolution by non-recursive filters) which can synthesize stable speech and it was utilized in speech synthesis. However, its impulse responses are short, and it is reported that quality degradation occurs even for fluctuation of the pitch. Whereas in [13], Mel-Frequency Cepstral Coefficients (MFCCs) were replaced with wavelet parameters both in the training and speech synthesis based on Hidden Markov Model (HMM). Ribeiro and Clark in [14] were using wavelets for F0 modeling under the assumption that they can be meaningfully related to linguistic units. However, this assumption was shown not to be accurate. Time-scale representation based on Continuous Wavelet Analysis (CWT) has previously been applied for detection and quantifying prosodic events regarding word and syllable prominence [15]. It was also seen in [16] that the CWT decomposition may be used with discrete cosine transform to model each prosody scale at a supra-segmental level.

To achieve higher quality speech synthesis, we propose a system based on CWT capable of modeling speech features in different temporal scales. Continuous fundamental frequency (contF0), Maximum Voiced Frequency (MVF), and spectral envelope of the speech signal are analyzed and decomposed by CWT. This approach is different from our earlier research in [17]. Finally, we compare its accuracy with the WaveNet architecture's performance using six different types of speakers. Experimental results indicate that our framework outperforms the conventional one in quality of analysis-synthesized speech. In the remaining part of this paper, we describe the wavelet synthesizer in Section 2. In Section 3, the experiments and simulation results are summarized. Finally, the discussions and conclusions are presented in Sections 4 and 5, respectively.

## 2. Methods

### 2.1. Continuous Wavelet Transform

A wavelet is a short waveform with finite duration, whose average value is zero. The continuous wavelet transform (CWT) can describe the signal in various transformations of a mother wavelet. Scaling the mother wavelet, the transform can capture high frequencies if the wavelet is compressed, and low frequencies if it is stretched. The process is repeated by translating the mother wavelet. The CWT output is an $MxN$ matrix where $M$ is the number of scales and $N$ is the length of the signal. The CWT coefficient at scale $a$ and position $b$ is given by:

$$W(a,b) = \frac{1}{\sqrt{a}} \int_{-\infty}^{+\infty} f(x)\psi(\frac{x-b}{a})dx \quad (1)$$

where $x$ is the input signal, and $\psi$ is the mother wavelet. This work will consider a decomposition strategy using the CWT and the Mexican hat mother wavelet. A set of 10 components is defined, where each component is approximately one octave apart.

The original signal can be recovered from the wavelet representation by inverse transform using the double-integral form over all scales and locations, $a$ and $b$ (for the proof, see [18] [19]):

$$f(x) = \int_{-\infty}^{+\infty}\int_{-\infty}^{+\infty} \frac{W(a,b)}{a^2\sqrt{a}}\psi(\frac{b-x}{a})dxda \quad (2)$$

Then we can obtain an approximation to the original signal by summing the scaled CWT coefficients over all scales

$$f(x) = \sum_{i=1}^{10} W_i(a,b;x) + \epsilon(x) \quad (3)$$

where $\epsilon(x)$ is the reconstruction error. CWT can analyze a speech waveform with a time-frequency resolution different from the Short time Fourier Transform (STFT). Figure 1 shows an STFT spectrogram and a CWT scalogram. A major difference between them is that the frequency bands in STFT have a fixed width, whereas in the CWT the frequency bands grow, shrink, scaled, and shifted to correlate with the signal's anomalies or events. This leads to high-frequency resolution with CWT at low frequencies and high time resolution at high frequencies.

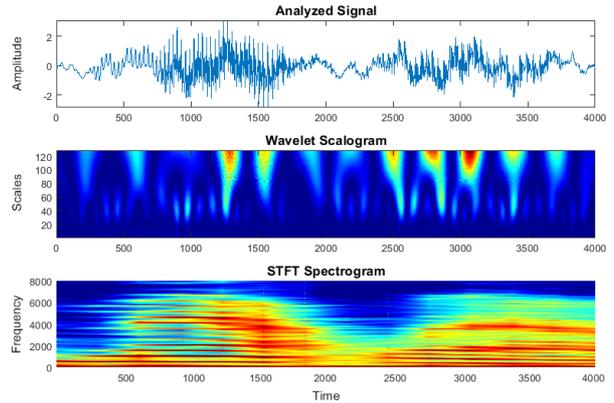

Figure 1: *Comparison of different time-frequency representations. Top Row: Original speech signal. Second row: Wavelet Scalogram using a Mexican hat wavelet $\psi(t)$, which is also a particular case of the m-th order derivative of a Gaussian wavelet. Third row: Short Time Fourier Transform spectrogram.*

### 2.2. Speech Analysis-Synthesis Framework

This work develops a statistical wavelet method of speech parameterization for speech synthesis. As shown in Figure 2, our framework consists of speech analysis and synthesis. During the analysis phase, as in our previous work in SPSS which was successfully used with a deep neural network-based TTS [20], continuous fundamental frequency (contF0) is calculated on the input waveform using a simple continuous pitch tracker [21].

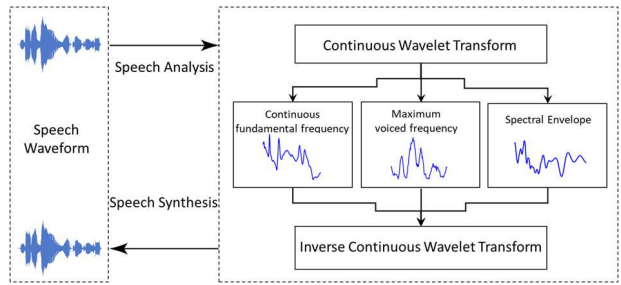

Figure 2: *Overview of proposed analysis-synthesis framework using CWT-based approximation of speech features.*

It should be noted that the contF0 algorithm may lead to erroneous tracking whenever the harmonic-to-noise ratio is low. To mitigate this issue, CWT coefficients were extracted from the speech signal and used to form a contF0 vector. The impact of this approach on contF0 performance is illustrated in Figure 3. Obviously, the contF0 obtained by CWT almost matches the reference pitch contour (that of REPEAR [22]) much better than baseline. It can also be seen here that the modified contF0 in the unvoiced region (frames from 135 to 158) is significantly much smaller than for the baseline.

Another excitation parameter is the maximum voiced frequency (MVF) which exploits both amplitude and phase spectra that are integrated into a maximum likelihood criterion to derive the MVF decisions [23]. Here, MVF is also decomposed by CWT. Similarly, we also approximate the extracted continuous spectral envelope [24] with the wavelet transform.

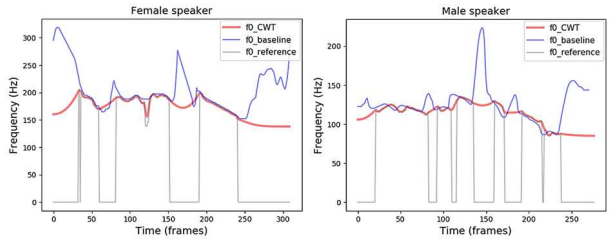

Figure 3: *Examples from female and male speakers of F0 trajectories estimated by the baseline (blue) and ground truth (grey) plotted along with proposed refined contF0 calculated from wavelet coefficients (red).*

During the synthesis phase, voiced excitation is made of principal component analysis (PCA) residuals overlap-added pitch synchronously. This voiced excitation is lowpass filtered frame by frame at the frequency given by the MVF parameter. In the frequencies higher than the actual value of MVF, white noise is applied. Voiced and unvoiced excitation is combined together, and the MGLSA (Mel-Generalized Log Spectrum Approximation) filter is used to synthesize speech.

The continuous wavelet vocoder has the apparent advantage of avoiding voicing decision per frame that may be considered to reduce the perceptual degradation caused by voicing decision errors. Moreover, it uses only two one-dimensional parameters for modeling the excitation, which is computationally feasible in deep neural network-based text-to-speech [25].

## 3. Experimental Evaluations

### 3.1. Experimental Conditions

In order to carry out the evaluation of the proposed system, we used six speakers from the CMU-ARCTIC database [26]; SLT, CLB, AWB, JMK, RMS, and KSP for evaluation, where SLT and CLB are female and others are male. The sampling frequency is set to 16 kHz with 16-bit linear quantization. The total number of utterances is 1132 per speaker, and the total utterance duration is about 1 hour per speaker. Acoustic features were extracted every 5ms after applying a window of 25ms.

In the experiment, we compare our proposed model with the following speech methods:

- **WaveNet:** It was trained by using 80-dim log-Mel spectrograms. The network architecture of the WaveNet was the same as that used in [27]. The total number of utterances is 6580 for training and 350 for testing (that is about 6 hours of recorded speech).
- **WORLD:** It was also used as a high-quality signal-processing-based vocoder [28]. Spectral envelopes and aperiodicity measurements obtained by utilizing the WORLD were converted to 59-dim Mel-cepstrum and 21-dim band aperiodicity. The total number of dimensions of a WORLD vocoder were 82 (59 Mel-cepstrum + 1 voiced/unvoiced flag + 1 F0 + 21 band aperiodicity).
- **Continuous** [17]**:** It was used as a baseline vocoder in this work. The total number of dimensions of the continuous vocoder were 26 (24 Mel-cepstrum + 1 MVF + 1 contF0).
- **Anchor:** It is a simple pulse-noise excitation vocoder [29]. This model was just used for the listening test.

### 3.2. Objective Results

In order to confirm whether the proposed model can reproduce the characteristics of the original speech, we evaluated spectral and fundamental frequency distortions between the natural speech and synthesized speech.

For cepstrum, the following Mel-Cepstrum Distortion (MCD) was applied:

$$MCD = \frac{10}{\log 10} \sqrt{\sum_{m=1}^{M} \left(c_{org}(m) - c_{synth}(m)\right)^2} \quad (4)$$

where $c_{org}$ and $c_{synth}$ are mel-cepstrum from original and synthesized speech, respectively, and $M$ is the order of mel-cepstrum. For F0, the following RMSE were applied:

$$RMSE = \sqrt{\frac{1}{N} \sum_{i=1}^{N} \left(F0_{org}^i - F0_{synth}^i\right)^2} \quad (5)$$

where $F0_{org}$ and $F0_{synth}$ denote the real and the synthesized continuous F0 features, respectively. The above MCD and RMSE were calculated for each frame and averaged over all the frames. Averages for male and female speakers have similarly been separated from each other. A lower MCD and F0-RMSE value indicate smaller distortion or prediction error.

The average MCD results from the natural and synthesized speech are presented in Table 1. Comparing the proposed method with the baseline and WaveNet, the suggested model decreases the value of MCD, proving that the continuous wavelet transform has a significant impact on the spectral feature of speech synthesis. It can also be seen that MCD of the proposed method was improved at almost the same level as the MCD of WORLD, which means that it could reproduce the original spectrum correctly.

Next, from Table 2, the root mean square error patterns are similar to the previous paragraphs' correlation results. It can be noticed that WORLD and the proposed vocoder could reproduce the original F0 with a relatively higher accuracy than WaveNet and the baseline vocoder. Thus, it was demonstrated that the proposed method could capture the voiced and spectral information with relatively higher accuracy and outperformed the state-of-the-art WaveNet vocoder.

Table 1: *Comparison of mel-cepstrum distortion between spectral features of natural speech and synthesized speech.*

| MCD (dB) | Male | Female |
|---|---|---|
| Baseline | 4.03 | 4.13 |
| WaveNet | 4.74 | 4.97 |
| WORLD | 3.31 | 3.27 |
| Proposed | 3.47 | 3.42 |

Table 2: *Comparison of F0-RMSE: Our framework versus the three vocoders.*

| RMSE (dB) | Male | Female |
|---|---|---|
| Baseline | 4.37 | 4.31 |
| WaveNet | 4.14 | 4.67 |
| WORLD | 3.42 | 3.51 |
| Proposed | 3.85 | 3.98 |

### 3.3. Subjective Results

In order to calculate the perceptual quality of the developed method, we performed a web-based MUSHRA (MUlti-Stimulus test with Hidden Reference and Anchor) listening experiment [30]. We evaluated natural sentences with the synthesized ones from the baseline, proposed, WORLD, WaveNet, and an anchor system. The participants had to assess the naturalness of each stimulus relative to the reference (which was the natural sentence), from 0 (highly unnatural) to 100 (highly natural). The listening test samples are available online[1]. Ten participants (7 males, 3 females) with a mean age of 33 years and no known hearing defects were invited to run the online perceptual test. On average, the MUSHRA test took 10 minutes.

Results are presented in Figure 4. The error bar represents 95% confidence interval. The proposed system is preferred over the baseline, but no significant differences are seen when compared against the WORLD system. Moreover, the proposed vocoder achieves higher naturalness than the WaveNet vocoder. Hence, our method presents a good alternative approach to other methods for the reconstruction of speech.

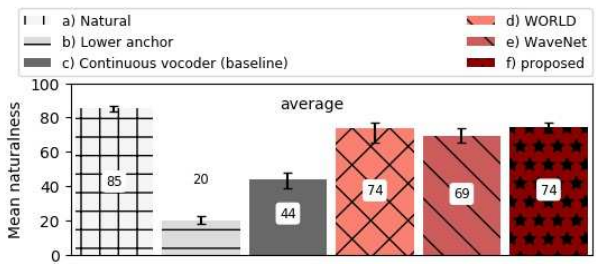

Figure 4: *Sound quality of synthesized speech.*

## 4. Discussion

We demonstrate that the wavelet transform provides a flexible approach to extract speech parameters and allows a multiscale decomposition of a speech signal. CWT is found to be significantly better than the standard STFT method. The continuous wavelet decomposition gives a possibility of adapting the weights of particular scales prior to reconstruction. This can be beneficial in the adaptation of speaking style.

Overall, the results suggest that the proposed method can compensate for the degradation of the acoustic features in the baseline and WaveNet models for both male and female speakers. Although the differences between the proposed and WORLD methods are relatively small (did not differ significantly), the wavelet-based method requires fewer acoustic parameters.

It is worth mentioning that although the original WaveNet paper describes that the WaveNet for TTS was trained by using speech data over 40 hours and successfully synthesized high-quality speech, this is only valid for the corpus from the CSTR voice cloning toolkit (VCTK) [31]. Oord et al. (2016) found that WaveNet conditioned on linguistic features could synthesize speech samples with natural segmental quality, but it had unnatural prosody by stressing wrong words in a sentence. In this paper, we also found that the WaveNet model did not perform well with CMU-ARCTIC corpus after testing it by six different speakers. Even though different accents and varying reading proficiency will impact the speech synthesis quality, this requires further investigation and research by those who plan to adopt the WaveNet system into their applications.

Additionally, the baseline contF0 contour was compared to the contF0 contour estimated with CWT by calculating the root mean square error for each test utterance. As expected, results showed significantly better performance for the wavelet method than for the baseline for all speakers. In the MUSHRA evaluation, the proposed method produces speech which sounds more persistent and more intelligible. But no significant differences were found when comparing it to the WORLD vocoder. Still, our system is more straightforward, i.e. uses fewer parameters.

## 5. Conclusions

The current paper has proposed a speech analysis-synthesis system based on wavelet decomposition. Our framework decomposes a multi-level representation of contF0, MVF, and spectral envelope using the continuous wavelet transform. Our work was supported by objective metrics of intelligibility and sound quality as well as subjective listening tests. We confirmed through experiments that our speech synthesis system was able to generate a natural-sounding synthetic speech and superior to state-of-the-art WaveNet vocoder on the CMU-ARCTIC database.

Future work includes incorporating deep neural networks to improve the synthetic speech quality and apply this technique to voice conversion. It also worth investigating post-filtering algorithms that are appropriate for our proposed method in the future.

## 6. Acknowledgements

The research was partly supported by the European Union's Horizon 2020 research and innovation programme under grant agreement No. 825619 (AI4EU), and by the National Research Development and Innovation Office of Hungary (FK 124584 and PD 127915). The Titan X GPU used was donated by NVIDIA Corporation. We would like to thank the subjects for participating in the listening test.

---

[1] https://malradhi.github.io/cwt_vocoder/